\documentclass[]{interact}

\usepackage{epstopdf}% To incorporate .eps illustrations using PDFLaTeX, etc.
\usepackage[caption=false]{subfig}% Support for small, `sub' figures and tables

\usepackage{natbib}% Citation support using natbib.sty
\bibpunct[, ]{(}{)}{;}{a}{}{,}% Citation support using natbib.sty
\renewcommand\bibfont{\fontsize{10}{12}\selectfont}% Bibliography support using natbib.sty

\usepackage{multirow}

\theoremstyle{plain}% Theorem-like structures provided by amsthm.sty

\theoremstyle{definition}

\theoremstyle{remark}

\begin{document}

%\articletype{ARTICLE}% Specify the article type or omit as appropriate

\title{Depth Estimation from Single-shot Monocular Endoscope Image Using Image Domain Adaptation And Edge-Aware Depth Estimation}

\author{
\name{
Masahiro Oda\textsuperscript{a,b}\thanks{CONTACT Masahiro Oda. Email: moda@mori.m.is.nagoya-u.ac.jp}, Hayato Itoh\textsuperscript{b}, Kiyohito Tanaka\textsuperscript{c}, Hirotsugu Takabatake\textsuperscript{d},\\ Masaki Mori\textsuperscript{e}, Hiroshi Natori\textsuperscript{f} and Kensaku Mori\textsuperscript{b,a,g}}
\affil{\textsuperscript{a}Information and Communications, Nagoya University, Nagoya, Japan; \textsuperscript{b}Graduate School of Informatics, Nagoya University, Nagoya, Japan; \textsuperscript{c}Department of Gastroenterology, Kyoto Second Red Cross Hospital, Kyoto, Japan; \textsuperscript{d} Department of Respiratory Medicine, Sapporo-Minami-Sanjo Hospital, Sapporo, Japan; \textsuperscript{e}Department of Respiratory Medicine, Sapporo-Kosei General Hospital, Sapporo, Japan; \textsuperscript{f}Department of Respiratory Medicine, Keiwakai Nishioka Hospital, Sapporo, Japan; \textsuperscript{g}Research Center for Medical Bigdata, National Institute of Informatics, Tokyo, Japan}
}
%\name{
%A.~N. Author\textsuperscript{a}\thanks{CONTACT A.~N. Author. Email: latex.helpdesk@tandf.co.uk} and John Smith\textsuperscript{b}}
%\affil{\textsuperscript{a}Taylor \& Francis, 4 Park Square, Milton Park, Abingdon, UK; \textsuperscript{b}Institut f\"{u}r Informatik, Albert-Ludwigs-Universit\"{a}t, Freiburg, Germany}
%}

\maketitle

\begin{abstract}
This paper proposes a depth estimation method from a single-shot monocular endoscopic image.
Automated understanding of endoscopic images is important for diagnosis and treatment assistance.
Not only the images themselves but also depth information about the images help make the understanding of endoscopic images, such as lesion-size measurements, more accurate.
Previous depth estimation methods have used stereo cameras or time-series images.
However, many endoscope imaging systems do not support the use of stereo endoscopes and video capturing.
Also, automatic classification or recognition of large number of previously stored single-shot monocular endoscopic images is required to perform retrospective studies of endoscopic image analysis.
We propose a depth estimation method from a single-shot monocular endoscopic image using Lambertian surface translation by domain adaptation and depth estimation using multi-scale edge loss.
The difficulty of the depth estimation is that we cannot obtain real endoscopic images and their corresponding depth images. Depth sensors cannot be attached to endoscopes because of the size limitation.
To tackle the difficulty, we employ a two-step estimation process including Lambertian surface translation from unpaired data and depth estimation.
The texture and specular reflection on the surface of an organ reduce the accuracy of depth estimations.
We apply Lambertian surface translation to an endoscopic image to remove these texture and reflections.
Then, we estimate the depth by using a fully convolutional network (FCN).
During the training of the FCN, improvement of the object edge similarity between an estimated image and a ground truth depth image is important for getting better results.
We introduced a muti-scale edge loss function to improve the accuracy of depth estimation.
We quantitatively evaluated the proposed method using real colonoscopic images.
The estimated depth values were proportional to the real depth values.
Furthermore, we applied the estimated depth images to automated anatomical location identification of colonoscopic images using a convolutional neural network.
The identification accuracy of the network improved from 69.2\% to 74.1\% by using the estimated depth images.
\end{abstract}

\begin{keywords}
Depth estimation; single-shot monocular endoscopic image; Lambertian surface translation
%; multi-scale edge loss
%Sections; lists; figures; tables; mathematics; fonts; references; appendices
\end{keywords}

\section{Introduction}

%Endoscopic diagnosis and treatment are commonly performed in medical institutions.
%An endoscope is inserted into a patient's body through an incised part or a natural orifice to capture interior views of the body.
%Endoscopes such as colonoscopes, gastroendoscopes, and bronchoscopes are used to diagnose aliments in digestive organs, trachea, and bronchi, respectively.
%Laparoscopes and thoracoscopes are used in surgery.
%However, diagnosis and treatment using these endoscopes can encounter some common problems.
%One problem is the difficulty of understanding 3D structures.
Diagnosis and treatment using endoscopes can encounter some common problem that the difficulty of understanding 3D structures.
%Most endoscopes capture images using a monocular camera.
Although some laparoscopes are equipped with stereo scope cameras, most endoscopes, including colonoscopes, have monocular cameras because of their small size.
%As a result, 3D information about an observation target is indeterminate in images from such cameras.
%Another problem is the limited of view.
%In endoscopic surgery, organs can be injured by the endoscope or other surgical tools. When such injuries occur, they do so outside of the field of view the endoscopes.
Image-based assistance by computers can help solve the problem.
%Recent progress in image-processing and -understanding technologies has the potential to provide assistance or navigation information in endoscopic diagnosis and treatment.
Examples of image-based endoscope assistance include endoscope navigation \citep{Hayashi16}, tracking \citep{Luo15}, lesion detection \citep{Yuan18,Brandao17}, and scene understanding during surgery \citep{Twinanda17,Aksamentov17}.
However, the lack of 3D information in monocular endoscopic images makes using such assistance difficult.
%For example, the evaluation of lesion size is difficult without knowing the distance from the camera to the lesion.
Therefore, reconstruction of 3D structures or depth estimation from endoscopic images is needed in many endoscope assistance applications.

Previous research has proposed 3D structure reconstruction or depth estimation from endoscopic images, including {\it Shape from Shading} and feature-point matching techniques \citep{Mair-Hein13}.
Because sensors for depth measurement cannot be used in combination with endoscopes due to the limitation of the size, image-based estimation is commonly performed to estimate depth.
Depth estimation from endoscopic images is commonly performed.
However, these approaches easily fail to estimate depths from real endoscopic images.
This is because endoscopic images can portray a huge variety of organ-surface textures.
%Such variety makes the calculation of {\it shape from shading} difficult.
Furthermore, organs appearing in endoscopic images show non-rigid deformations.
Such deformations decrease the matching accuracies of feature points in images.
Thus, a new depth estimation method that does not rely on these previous techniques is needed.
Recently, many deep learning-based depth estimation methods from indoor or driving images \citep{Godard17,Luo18cvpr,Prasad19,Liu19cvpr,Guo19,Ma19,Zhang19,Chabra19,Ren19} and endoscopic images \citep{Visentini17,Mahmood17,Mahmood18-1,Mahmood18-2,Rau19,Liu18,Liu19,Luo19} have been proposed.
Among the estimation methods from endoscopic images, Visentini-Scarzanella et al. \citep{Visentini17} trained and tested only using images taken from a phantom.
Mahmood et al. \citep{Mahmood17,Mahmood18-1,Mahmood18-2} performed quantitative evaluation on a phantom and a porcine colon datasets.
Rau et al. \citep{Rau19} also evaluated their method on a phantom dataset.
Quantitative performances of the previous methods on real human dataset remain obscure.
Liu et al. \citep{Liu18,Liu19} and Luo et al. \citep{Luo19} performed estimations from time-series and stereo images, respectively.
Depth estimation from a single-shot image is still challenging task as a baseline of 
depth estimation methods using time-series or stereo camera images.
The single-shot image-based depth estimation is useful to perform automatic classification or recognition of large number of previously stored single-shot monocular endoscopic images in retrospective studies of endoscopic image analysis.

We propose a depth estimation method from a single-shot monocular endoscopic image.
The difficulty of the depth estimation is that we cannot obtain real endoscopic images and their corresponding depth images. Depth sensors cannot be attached to endoscopes because of the size limitation.
To tackle the difficulty, we employ a two step estimation process including Lambertian surface translation from unpaired data and depth estimation.
%Because an endoscopic image and its corresponding depth image cannot be obtained due to the size limitation of endoscopes, we employ a two step estimation process including Lambertian surface translation from unpaired data and depth estimation.
While depth estimation 
%or 3D reconstruction 
from a Lambertian surface in a 2D image is possible, light reflections on organ surfaces in real endoscopic images contain not only diffuse but also specular reflections.
Also, textures on organ surfaces make depth estimation difficult.
We remove such specular reflections and textures on the organ surface by using a real to Lambertian surface translation by a domain adaptation technique.
The domain adaptation translation is performed by a fully convolutional network (FCN), which is trained in an unpaired image training framework.
% to use real images in training.
Then, the domain translated images are processed by a depth estimation network.
We use a DenseNet-based encoder-decoder-style FCN as the depth estimation network.
We propose a multi-scale edge loss that helps to give highly accurate depth estimations.
Object edge information is important to evaluate the quality of depth estimations.
%The multi-scale edge loss evaluates the estimation quality of object edges, including sharp to mild shifts.
The multi-scale edge loss evaluates the estimation quality of object edges, including clear to blurred edges.
We obtain results from the depth estimation network trained using the multi-scale edge loss. %消す？
We performed quantitative evaluation of depth estimation results on real human dataset that previous methods failed to report.

The contributions of this paper can be summarized as: (1) a Lambertian surface translation process by domain adaptation to improve depth estimation accuracy, (2) a multi-scale edge loss for FCN-based depth estimation, and (3) quantitative evaluation of depth estimation on real human dataset. Use of the loss improved the depth estimation accuracy.

\section{Depth Estimation Method}

\subsection{Overview}

\begin{figure}[tb]
\begin{center}
\includegraphics[width=0.98\textwidth]{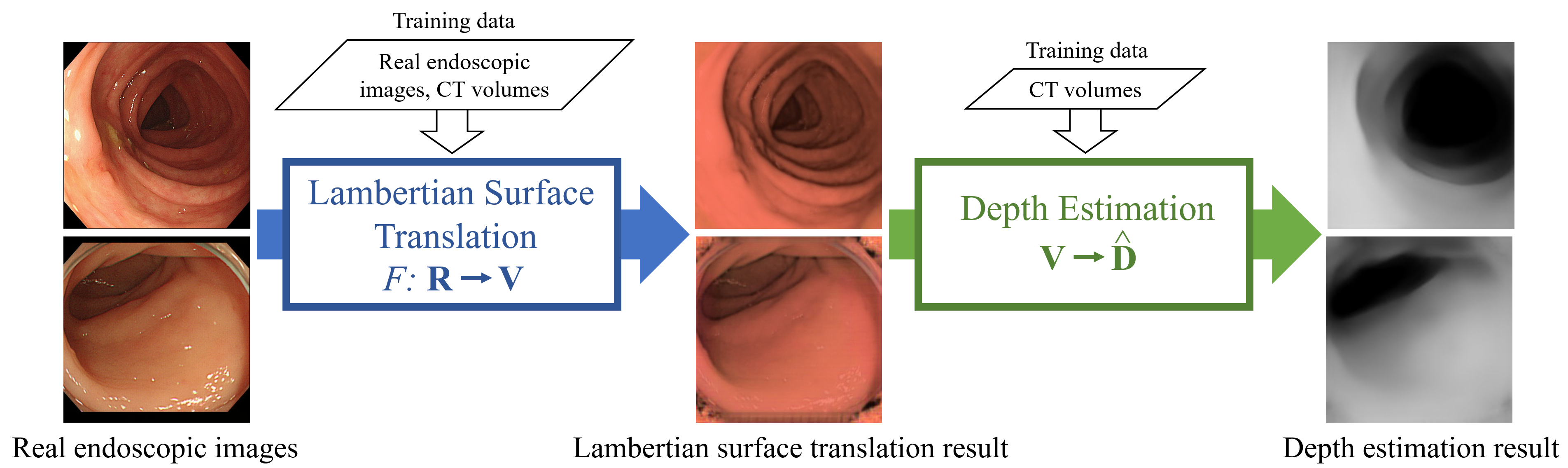}
\end{center}
\caption{Process flow of proposed depth estimation method. A single-shot monocular real endoscopic image is converted by the Lambertian surface translation process to remove specular light reflection and textures on organ surfaces. The depth estimation method trained using virtual endoscopic images and depth images generated from CT volumes is applied to the image to estimate a depth image.}
\label{fig:overview}
\end{figure}

A single-shot monocular real endoscopic image is the input of our method.
Lambertian surface translation is applied to the image to remove specular reflections and textures on the organ surface.
The translated image is processed by the depth estimation network.
The network is trained using the multi-scale edge loss.
The output of the network is an estimated depth image.
The process flow of the proposed method is shown in Fig. \ref{fig:overview}.

Depth images that correspond to real colonoscopic or bronchoscopic images are difficult to obtain because of the size limitation of such endoscopes.
Therefore, we employ an unpaired training framework to estimate depth images from real endoscopic images.
Our method establishes a depth estimation network from unpaired training data.

\subsection{Lambertian Surface Translation by Domain Adaptation}
%reduction of specular reflection and texture on surface by domain adaptation

Following the {\it shape from shading} theory \citep{Sfs}, we can estimate the shape of a 3D surface from a 2D image that captures a Lambertian surface having the homogeneous reflection property.
This is because surface normals are calculated from shading or intensity values on the surface in a perfect diffuse reflection.
The intensity value on the surface $\phi$ is calculated as 
\begin{equation}
\phi = \rho {\bf N} \cdot {\bf S} = \rho \cos \theta,
\end{equation}
where $\rho$ is the diffuse reflectance rate, ${\bf N}$ is the surface normal, ${\bf S}$ is the light direction, and $\theta$ is the angle between ${\bf N}$ and ${\bf S}$.
In endoscopic images, the light direction is the same as the camera direction because the light and the camera are mounted at almost the same position.
From the above equation, we obtain
\begin{equation}
\theta = \cos^{-1} \left( \frac{\phi}{\rho} \right),
\end{equation}
to estimate the surface normal.
The surface normal vector is calculated under the assumption that the surface is smooth and continuous.
Based on the above equations, the shape or depth of a 3D surface can be calculated from the surface normals if we have the $\rho$ of the surface.

However, light reflections on organ surfaces contain not only diffuse but also specular reflections.
%Specular reflections make estimation of the shape of a 3D surface from a 2D image difficult.
In the dichromatic reflection model \citep{Dichromatic}, light reflection on the surface is represented as the sum of the diffuse and specular components.
In a previous method \citep{Umeyama04}, parameters in the model were calculated from object images taken by a rotating polarizer.
This approach is difficult to apply to in-vivo endoscopic images because there is no commercially-available endoscopes including colonoscope or bronchoscope that has the function of rotating polarizing filters.
% cannot obtain polarized images.
Because it is quite difficult to measure the light reflection properties on the in vivo organ surfaces, including diffuse and specular reflectance rates that are necessary to calculate the dichromatic reflection model, an alternate approach to reduce the effects of specular reflections on shape estimation is necessary.
Also, there are organ-specific textures on the organ surfaces.
Such textures make shape estimation difficult.
%textureの問題について書く

%Specular light reflection and texture on organ surface make shape or depth estimation difficult.
We remove specular light reflection and textures on organ surfaces from endoscopic images by using a translation based on a domain adaptation technique from real to Lambertian surface domains.
%Organ surfaces in endoscopic images are translated to Lambertian surfaces using a translator built by the domain adaptation.
Because we build the translator using a data-driven approach, we can skip measurement of the light reflection properties on the in vivo organ surfaces.
We use an FCN to perform the domain adaptation.
%The FCN is trained to translate organ surfaces in endoscopic images from real to Lambertian surface domains.
Sets of endoscopic images in real and Lambertian surface domains are denoted as ${\bf R}$ and ${\bf V}$, respectively.
The translator performs mapping $F : {\bf R} \rightarrow {\bf V}$.
We train the FCN using CycleGAN \citep{Cyclegan}, which is a training framework that uses unpaired real and virtual endoscopic images.
Real endoscopic images ${\bf r}_{i} (i=1,\dots, I) \in {\bf R}$ are taken from patients during endoscopic diagnoses.
Endoscopic images purely containing Lambertian organ surfaces are generated as virtual endoscopic images.
Virtual endoscopic images ${\bf v}_{j} (j=1, \ldots, J) \in {\bf V}$ are generated from CT volumes of patients using a volume rendering technique \citep{Mori03} that uses diffuse reflectance as a light reflection model on organ surfaces.
$I$ and $J$ are the numbers of real and virtual endoscopic images, respectively.
The translator $F$ is implemented as a U-Net \citep{Unet} with instance normalization \citep{Instancenormalization} after each convolution layer.
%discriminatorの説明
%Examples of domain adaptation translation results by a trained $F$ are shown in Fig. \ref{fig:domaintranslationresult}.

%\begin{figure}[bt]
%\begin{center}
%\includegraphics[width=0.98\textwidth, clip, trim=0 220 40 0]{fig/domaintranslateresult.eps}
%\caption{Examples of Lambert surface translation by domain adaptation. Inputs to translator are real colonoscopic images shown in top row. Translated images are shown in bottom row.}
%\label{fig:domaintranslationresult}
%\end{center}
%\end{figure}

\subsection{Depth Estimation Network}

\subsubsection{Network Structure}
Depth estimation from a monocular single-shot 2D image is an ill-posed problem.
However, with the development of deep learning-based depth estimation techniques, reasonable depths can be estimated from such images.
Among many network structures for depth estimation, the encoder-decoder-style depth estimation FCN \citep{Alhasim19} based on the DenseNet-169 \citep{Densenet} produces accurate and high-resolution depth estimations.
The depth estimation FCN has a pre-trained DenseNet-169 as an encoder.
The decoder of the depth estimation FCN consists of upsampling and convolution layers.
%ネットワーク構造図で示す？
Feature maps of many resolutions in the encoder are sent to corresponding layers in the decoder by skip connections (concatenation operation).
The skip connections help keep spatial resolutions and produce detailed estimations of depth images.
DenseNet-169 in the encoder is pre-trained on the ImageNet.
Even though the weights on the network are pre-trained to classify images, the transfer learning from classification to depth estimation improves depth estimations.

\subsubsection{Multi-scale Edge Loss}
An appropriate loss function for depth estimation is needed to get better training results from the FCN.
L1 or L2 norms are commonly used as loss functions that evaluate the difference between the ground truth and estimated depth images.
%The similarity of object edges in depth images is also important for training a depth estimation network \citep{Alhasim19}.
Object edge information in depth images is also important for training a depth estimation network \citep{Alhasim19}.
Alhashim and Wonka \citep{Alhasim19} introduced first-order differential of depth values in the loss function to consider object edge difference.
%他にedgeをlossに使う文献挙げる
%However, they consider only small edges.
%They fail to consider large edges of objects in their loss function.
They calculated the first-order differential of depth values in a small local region, such as a region in $3\times3$ pixels.
However, the use of a small local region results in consideration of very clear edges in the loss function.
%However, the use of first-order differential reflects only very sharp edges to their loss.
In some cases, edges of objects in endoscopic depth images are not clear because endoscopic images are blurred when movement of the endoscope tip was quick.
%In many cases, edges of objects have mild shifts of depth values.
%Such mild edges can be found frequently in endoscopic depth images because the shapes of organs are smooth.
Because the application target of the method proposed by Alhashim and Wonka \citep{Alhasim19} is made up of natural images, their loss function is not suitable for endoscopic images.

%項をlossと呼ぶか式全体をlossと呼ぶか
We propose a multi-scale edge loss term in a loss function that takes clear and blurred object edges into account.
The loss evaluates the difference of the edges in depth images that have multiple thickness.
This term is effective in quality evaluations of endoscopic depth images.
We represent a ground truth depth image as ${\bf D}$ and an estimated depth image as ${\bf \hat{D}}$.
Our loss function $L$ is represented as
\begin{equation}
L({\bf D}, {\bf \hat{D}})=\lambda L_{d}({\bf D}, {\bf \hat{D}}) + L_{S}({\bf D}, {\bf \hat{D}}) + L_{e}({\bf D}, {\bf \hat{D}}),
\end{equation}
where $L_{d}({\bf D}, {\bf \hat{D}})$ is the point-wise L1 loss term and $L_{S}({\bf D}, {\bf \hat{D}})$ is the structural similarity (SSIM)-based loss term.
Definitions of the terms can be found in Alhashim and Wonka \citep{Alhasim19}.
$L_{e}({\bf D}, {\bf \hat{D}})$ is the multi-scale edge loss term, which is described as
\begin{equation}
L_{e}({\bf D}, {\bf \hat{D}}) = \frac{1}{P}\sum_{p}^{P} \max \{ | {\bf \hat{G}}^{(3)}_{p} - {\bf G}^{(3)}_{p} |, \ | {\bf \hat{G}}^{(5)}_{p} - {\bf G}^{(5)}_{p} |, \ | {\bf \hat{G}}^{(7)}_{p} - {\bf G}^{(7)}_{p} | \},
\end{equation}
where $p$ is the index of a pixel in a depth image and $P$ is the total number of pixels in a depth image.
${\bf G}^{(3)}$ is the edge image obtained by applying the $3\times3$ differential filter to ${\bf D}$.
Similarly, ${\bf G}^{(5)}$ and ${\bf G}^{(7)}$ are obtained by applying the $5\times5$ and $7\times7$ differential filters to ${\bf D}$, respectively.
The $3\times3$, $5\times5$, and $7\times7$ differential filters calculate differentials between adjacent pixels, pixels at one pixel intervals, and pixels at two pixel intervals, respectively, in ${\rm D}$.
${\bf \hat{G}}^{(3)}, {\bf \hat{G}}^{(5)}$, and ${\bf \hat{G}}^{(7)}$ are obtained by applying the three scales differential filters to ${\bf \hat{D}}$.
%The multi-scale edge loss term evaluates the difference of the edges between the ground truth and an estimated depth image. %消す？
%The edges are calculated as the maximum values of differentials in the three scales on a depth image. %消す？

\subsubsection{Network Training}
We train the depth estimation FCN using the multi-scale edge loss.
Virtual endoscopic images ${\bf v}_{j}$ and their corresponding depth images ${\bf d}_{j}$ are fed to the FCN.
Depth images ${\bf d}_{j}$, which correspond to ${\bf v}_{j}$, are generated from the CT volumes of patients.
The depth images have grayscale intensity values that correspond to the distance from a virtual camera position to a position on the surface of an inner wall of a hollow organ.

\subsection{Depth Estimation}
To estimate a depth image from a real endoscopic image, two trained FCNs are used.
A real endoscopic image is processed by $F$ to translate into a Lambert surface.
The translated image is then processed by the depth estimation FCN to obtain a depth estimation result ${\bf \hat{d}}$.

We perform a simple correction process of depth values of the depth estimation result.
The correction process contains scaling and translation of depth values.
The correction process is applied to each depth value in ${\bf \hat{d}}$ by
\begin{equation}
\tilde{d}_{k} = s\hat{d}_{k} + t,
\end{equation}
where $s$ and $t$ are the scaling and translation coefficients of the correction process.
$k$ is an index of pixels in ${\bf \hat{d}}$.
$\hat{d}_{k}$ is a depth value of $k$-th pixel in ${\bf \hat{d}}$.
$\tilde{d}_{k}$ is a corrected depth value of $k$-th pixel in a corrected depth image ${\bf \tilde{d}}$.
We obtain the corrected depth image ${\bf \tilde{d}}$ as the final depth estimation result.

\section{Experiments and Results}
%domain adaptation translator, Lambert domain translator, Lambert surface domain translation
We evaluated the proposed method quantitatively.
We applied the method to real colonoscopic images with a point depth to evaluate the accuracy of the depth estimation.
Also, to evaluate the usefulness of the estimated depth images in automated endoscopic scene understanding, we performed automated anatomical location identification of colonoscopic images using a convolutional neural network (CNN).
The estimated depth images were used for image classification.

We generated $J=8,085$ virtual colonoscopic images and corresponding depth images from six cases of colon CT volumes.
These images were taken during manual fly-through in the colon in the CT volumes.
For the training of the Lambertian surface translator, the virtual colonoscopic images and $I=13,406$ real colonoscopic images were used.
The generators and discriminators were trained in $400$ iterations with a $38$-minibatch size.
The depth estimation network was trained using $J=8,085$ pairs of virtual colonoscopic and depth images.
The training epoch was $7$, and the minibatch size was $10$.
The parameter value was set as $\lambda=0.1$.

In the correction process of depth values, values of the parameters were set as $s=0.73$ and $t=-3.0$.
These values were selected experimentally.

The size of all images used in our method was $256 \times 256$ pixels.
The virtual and real colonoscopic images were in color, and the depth images were grayscale.
The brightest and darkest intensity values in the depth images correspond to depth values of 0 and 100 mm, respectively.
%Examples of the Lambertian surface translation and the depth estimations are shown in Fig. \ref{fig:result_depthestimation}.

\subsection{Ablation Study and Comparative Study}\label{ssec:ablation}

We performed an ablation study of the proposed method. 
We proposed the Lambertian surface translation (LST) method and the depth estimation by the FCN using the multi-scale edge loss (ME loss) (\textit{LST + Depth estimation with ME loss} (Proposed)).
To confirm the effectiveness of using the multi-scale edge loss, we used the mean absolute error loss (MAE loss) as the loss function to train the depth estimation FCN (\textit{LST + Depth estimation with MAE loss}).
We also compare results obtained by using and without using the LST.
To perform the comparison, we made a depth estimation method from real colonoscopic images without using the LST (\textit{Depth estimation without LST}).
We made the method by using an image translation based on a domain adaptation technique trained using CycleGAN.
We need to use the unpaired training technique because the real colonoscopic and depth images are unpaired.

\begin{figure}[bt]
\begin{center}
\includegraphics[width=0.98\textwidth]{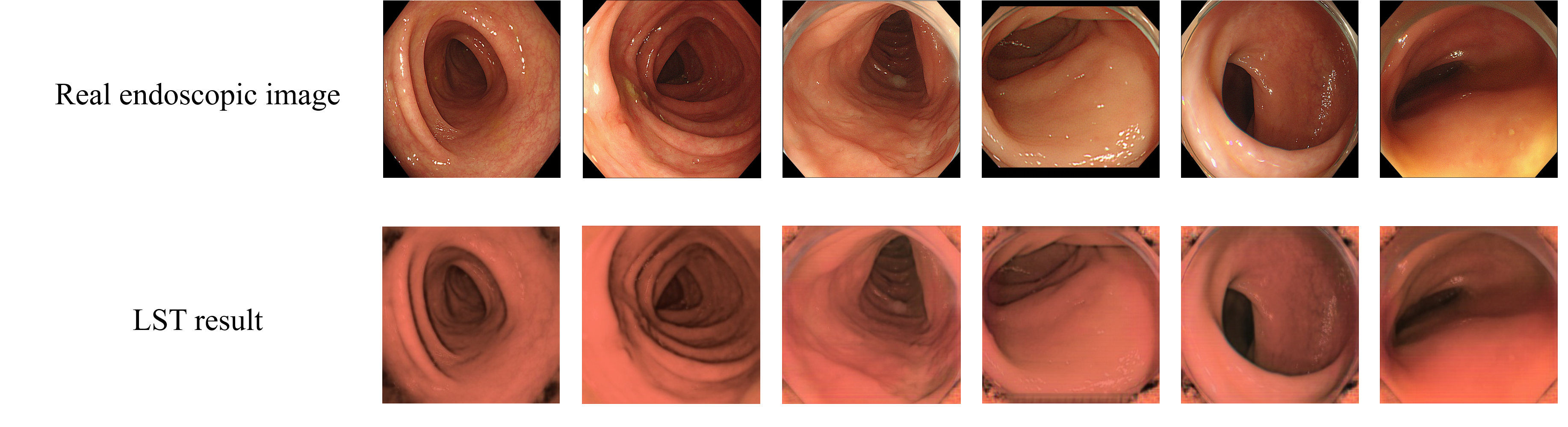}
\includegraphics[width=0.98\textwidth]{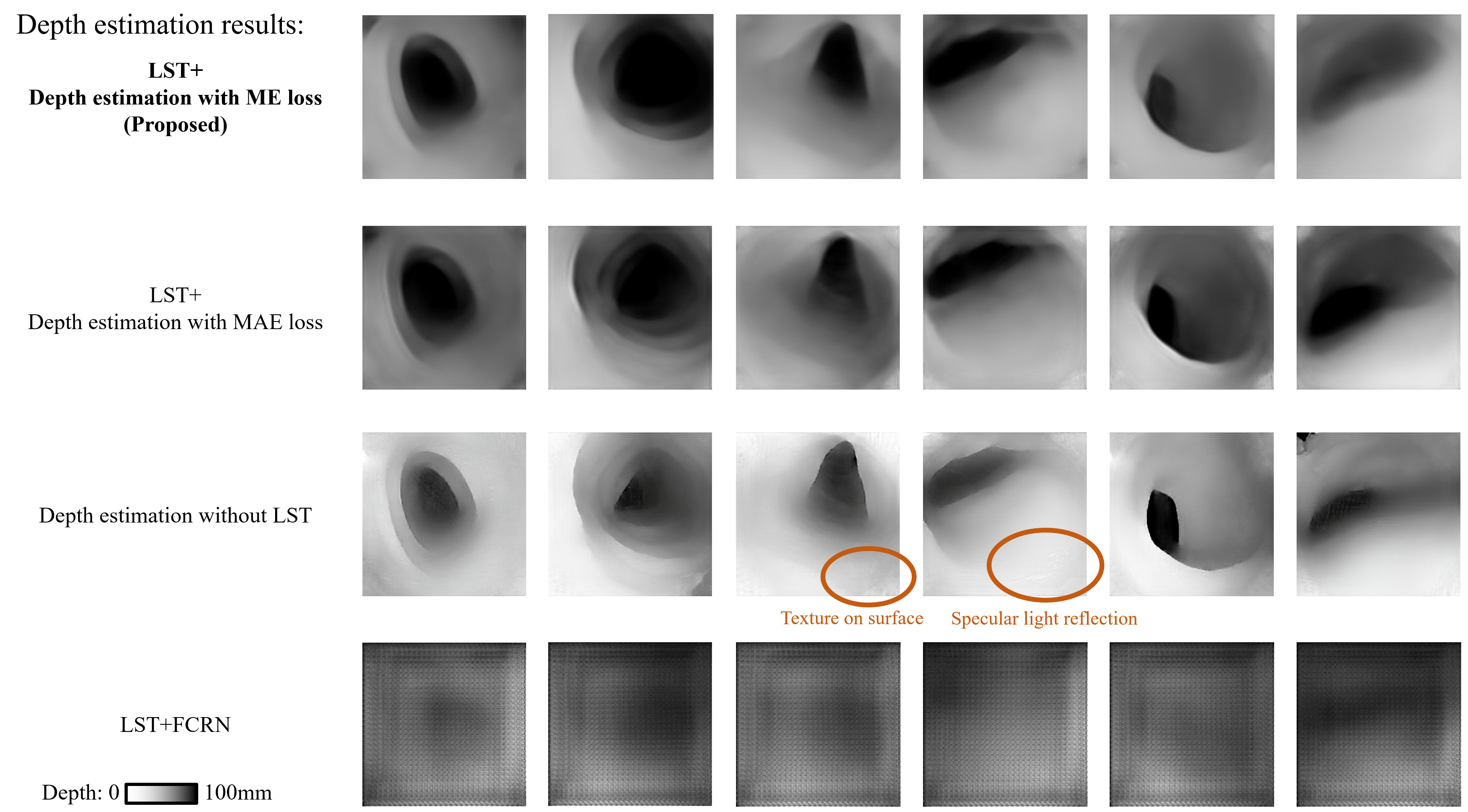}
\caption{Results of LST and depth estimations. The real colonoscopic images shown in the top row were used as the input of the translator. 
Results of LST are also shown.
Depth estimation results using proposed method are shown as \textit{LST + Depth estimation with ME loss}, which uses LST and depth estimation FCN trained using multi-scale edge loss (ME loss).
Results obtained using mean absolute error loss (MAE loss) are indicated as \textit{LST + Depth estimation with MAE loss}.
Results obtained without using LST are indicated by \textit{Depth estimation without LST}.
Results obtained by FCRN \citep{FCRN} are indicated as \textit{LST + FCRN}.
%The translated images shown in the middle row were then used as the input of the depth estimator.
}
\label{fig:result_depthestimation}
\end{center}
\end{figure}

Depth estimation results of the above three methods are shown in Fig. \ref{fig:result_depthestimation}.
The results of the proposed method and the \textit{LST + Depth estimation with MAE loss} represented the shape of the colonic surfaces accurately.
In the results of the \textit{Depth estimation without LST}, depth values were affected by texture and specular light reflections on the surface.
Estimated depth values in these areas were not accurate.

We compared the proposed method with a previously proposed depth estimation method from a single-shot monocular image \citep{FCRN}.
The previous method uses a fully convolutional residual network (FCRN) to estimate depth images.
We used the method to estimate depth images from the results of the LST (\textit{LST + FCRN}).
The FCRN was trained using 8,085 pairs of virtual colonoscopic and depth images.
The training epoch was 40, and the minibatch size was 16.
Depth estimation results of the \textit{LST + FCRN} were shown in Fig. \ref{fig:result_depthestimation}.
Colonic surface shapes were not represented in the results.

\subsection{Evaluation of Depth Estimation Accuracy}

%We prepared depth value known real colonoscopic images.
We used graduated endoscopy forceps for measurement of the sizes or lengths in the endoscopic images.
%The forceps displays millimeter-scaled labels along its bar-like body.
The forceps displays 2 mm-scaled labels along its bar-like body.
During colonoscope insertions into patients, we aligned the forceps from the camera position of the colonoscope to the colonic wall to measure depth values.
Then, we took real endoscopic images that included the aligned forceps.
The measured depth values were used as the ground truth of the depth values.
The image was called real endoscopic image with a point depth.

\begin{figure}[tb]
\begin{center}
\includegraphics[width=0.95\textwidth]{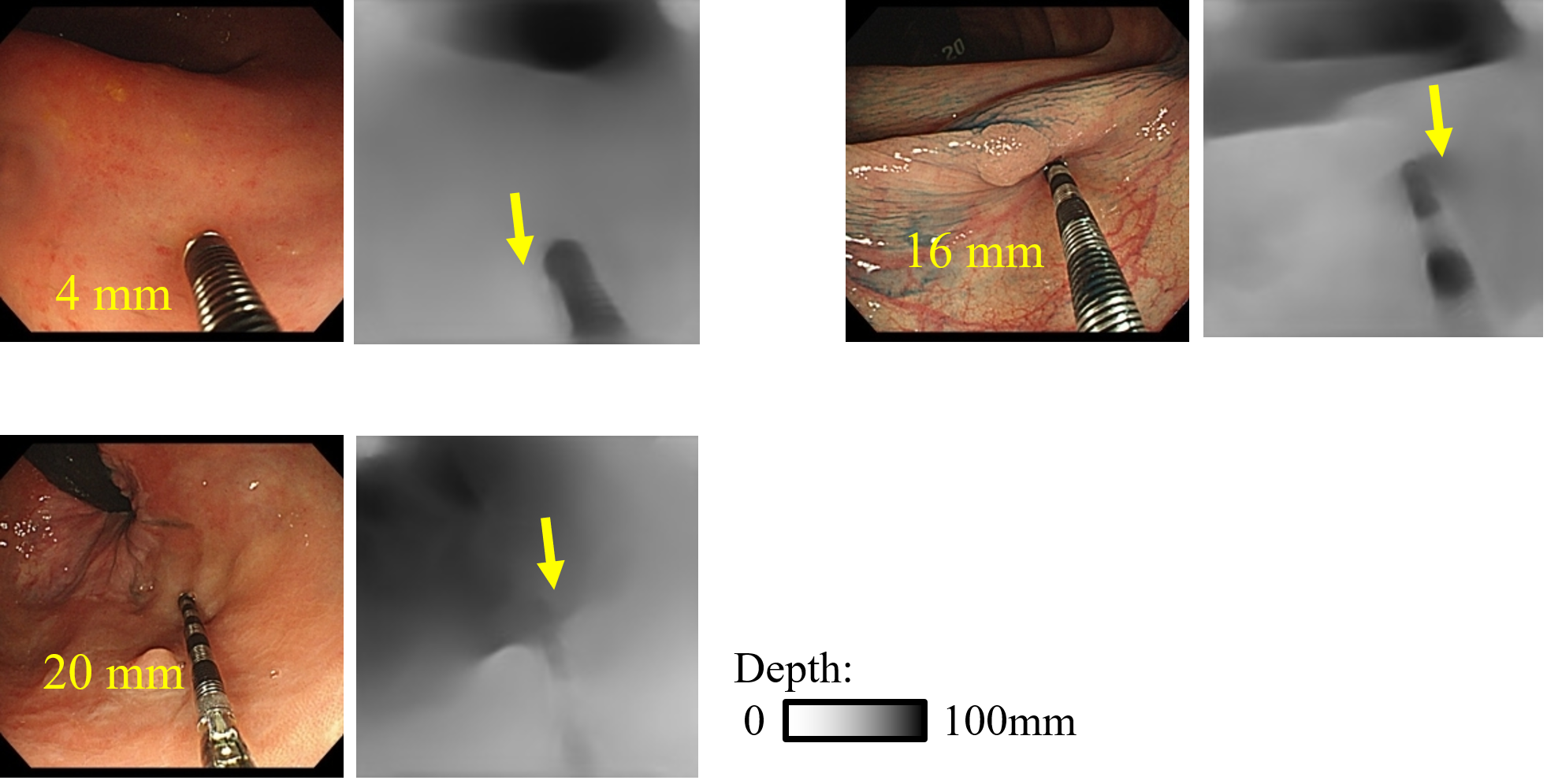}
\end{center}
\caption{Real endoscopic images with a point depth and depth estimations. Lengths of forceps are shown in real endoscopic images. 
%Estimated depth values 
Depth values estimated by proposed method were measured at the positions on the colonic wall near the location where the forceps pointed (indicated by arrows). Measuring estimated depth values on the forceps was avoided because estimated depth values on the forceps are not accurate.}
\label{fig:result_depthvalues_forceps}
\end{figure}

We applied the proposed method to 60 real endoscopic images with a point depth.
In the estimated depth images, we picked up an estimated depth value at a position on the colonic wall near the location where the forceps pointed as shown in Fig. \ref{fig:result_depthvalues_forceps}.
We avoided picking up an estimated depth value on the forceps because estimated depth value on the forceps is not accurate.
We selected an estimated depth value on the colonic surface near the forceps.
We compared the ground truth and estimated depth values.
The results are shown in Table \ref{tab:result_depthvalues}.
Even though absolute values of the estimated depth values are different from the ground truth, the averaged estimated depth values were clearly increase along with the ground truth.
%Even though absolute values of the estimated depth values are different from the ground truth, the averaged estimated depth values were clearly proportional to the ground truth.
The correlation coefficient of the ground truth and estimated depth values was 0.45.
It means the estimated depth values by the proposed method were correlated with the ground truth depth values.

%\begin{figure}[tb]
%\begin{center}
%\begin{tabular}{cc}
%\includegraphics[width=0.2\textwidth, clip, trim=0 300 740 0]{fig/result_depthknown_forceps.eps} & 
%\includegraphics[width=0.38\textwidth, clip, trim=0 300 570 0]{fig/result_depthknown_graph.eps} \\
%(a) & (b)
%\end{tabular}
%\end{center}
%\caption{(a) real endoscopic images with a point depth and depth estimation results. Estimated depth value was measured at positions where forceps pointing. (b) relationships between ground truth and estimated depth values.}
%\label{fig:result_depthvalues_forceps_graph}
%\end{figure}

\begin{table}[tb]
\begin{center}
\caption{Relationships between ground truth and estimated depth values.}
\label{tab:result_depthvalues}
\begin{tabular}{|c|c|c|} 
\hline
Ground truth (mm) & Average estimated (mm) & Number of images \\ \hline
4 and 6 & {\bf 5.38} & 5 \\ \hline
8 and 10 & {\bf 11.05} & 5 \\ \hline
12 and 14 & {\bf 15.73} & 11 \\ \hline
16 and 18 & {\bf 16.66} & 16 \\ \hline
20         & {\bf 20.21} & 23 \\ \hline
\end{tabular}
\end{center}
\end{table}

\subsection{Application of Depth Estimation Results to Anatomical Location Identification from Colonoscopic Images}

\begin{figure}[bt]
\begin{center}
\includegraphics[width=0.98\textwidth]{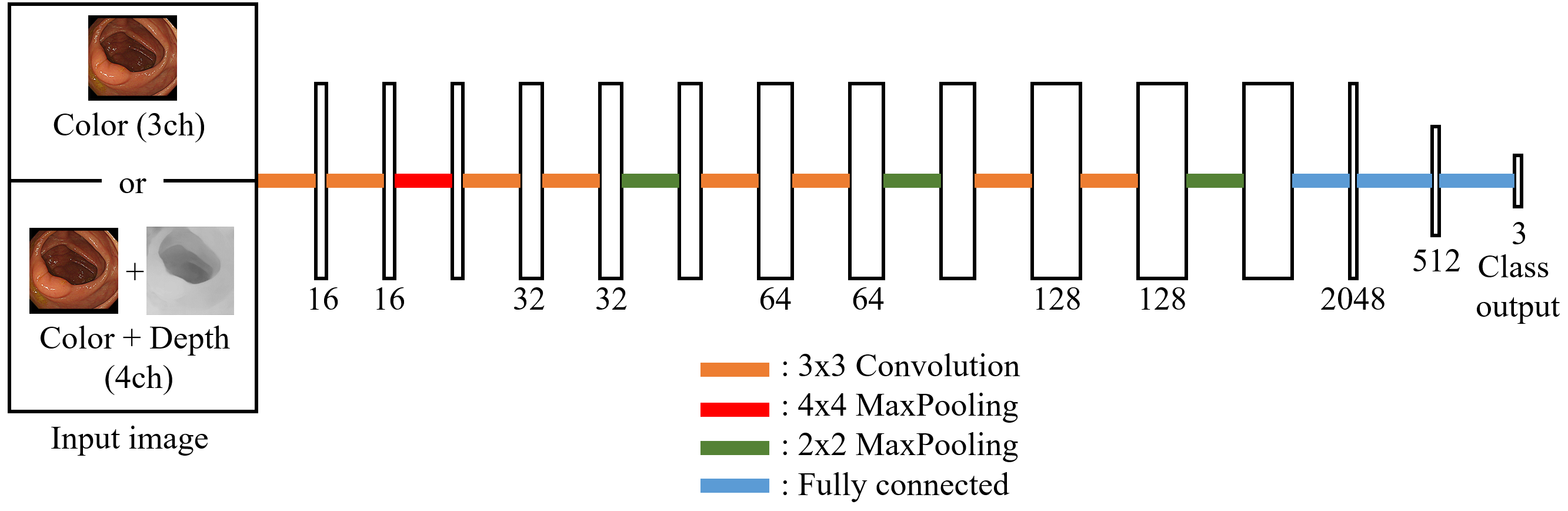}
\caption{Structure of CNN for anatomical location identification of colonoscopic images. The boxes are layers in the CNN. The numbers shown below the boxes are the numbers of kernels or neuron units. The three neurons in the final layer correspond to the three classes.}
\label{fig:classificationcnn}
\end{center}
\end{figure}

%scene-identification, location-identification
Depth information is useful in automated location-identification.
To evaluate the usefulness of our method for this identification, we used the estimated depth images in CNN-based anatomical location identification of real colonoscopic images.
We made a location identification CNN, as shown in Fig. \ref{fig:classificationcnn}.
The CNN classified an input image into three classes: the ileocecal area and ascending colon, the descending colon, and the rectum.
We trained the CNN using two sets of images: a set of only real colonoscopic images and a set of combined images of real colonoscopic images and depth images.
The combined images were made by combining the real colonoscopic images and depth images in the color channel.
%We trained the CNN using two sets of images: a set of only real colonoscopic images and a set of both real colonoscopic images and depth images, which combined in the color channel.
We used 2131 real colonoscopic images that were not used in training either the domain adaptation translator or the depth estimation network.
%Depth images were generated from them using the trained depth estimation network.
Depth images were generated from them using the four methods used in \ref{ssec:ablation}.
80\% and 20\% of the images were used for training and evaluation of the CNN, respectively.
Separation of images into the training and evaluation sets were performed randomly.
The CNN was trained in 50 epochs with a 50-minibatch size.
We compared the classification accuracies of the CNN when trained using the two sets of images.

We performed the random training/evaluation sets separations and CNN trainings five times.
The classification accuracies are shown in \ref{tab:result_cnn}.
From this table, the depth images contributed to improving the classification accuracies of CNN.
Among the four depth estimation methods, using the depth images generated by the proposed method resulted in obtaining the highest classification accuracy.
This shows that the depth images generated by the proposed method include much useful information for understanding real colonoscopic images.
%The classification accuracy of the location-identification CNN were \textcolor{blue}{69.2 $\pm$ 1.5 \%} when using only the real colonoscopic images as input and \textcolor{blue}{74.1 $\pm$ 2.5 \%} when using both the real colonoscopic images and depth images as input (averages \textcolor{blue}{and standard deviations} of five times 
%experiments
%of training and evaluation
%).
%The classification accuracy was improved by using the depth images.
%This shows that the depth images include much useful information for understanding real colonoscopic images.

\begin{table}[tb]
\begin{center}
\caption{Classification accuracies of location identification CNN. R and D mean real colonoscopic and depth images. Averages and standard deviations in five times experiments were shown.}
\label{tab:result_cnn}
\begin{tabular}{|c|c|}
\hline
Input of CNN & Classification accuracy (Ave. $\pm$ S.D.) \\ \hline
R & 69.2 $\pm$ 1.5 \% \\ \hline
\begin{tabular}{c}
R, D generated by \\ \textit{LST + Depth estimation with ME loss} \\ (Proposed)
\end{tabular}
 & \textbf{74.1 $\pm$ 2.5 \%} \\ \hline
\begin{tabular}{c}
R, D generated by \\ \textit{LST + Depth estimation with MAE loss}
\end{tabular}
 & 70.8 $\pm$ 2.1 \% \\ \hline
\begin{tabular}{c}
R, D generated by \\ \textit{Depth estimation without LST}
\end{tabular}
 & 69.9 $\pm$ 2.4 \% \\ \hline
\begin{tabular}{c}
R, D generated by \\ \textit{LST + FCRN} \citep{FCRN}
\end{tabular}
 & 63.7 $\pm$ 8.8 \% \\ \hline
\end{tabular}
\end{center}
\end{table}

\section{Discussion}

Machine learning-based depth estimation from colonoscopic images is difficult because commercially-available colonoscopes cannot obtain depth images.
By translating real colonoscopic images to virtual images by the LST, the depth estimation network trained in the virtual image domain is applicable for colonoscope depth estimation.

%The proposed depth estimation method estimated depth images from colonoscopic images.
In Fig. \ref{fig:result_depthestimation}, the depth images generated by the proposed method reflect the distance from the camera to the colonic wall at any point on the image.
Importance of the LST can be observed by comparing the results of the proposed method and the results obtained without using LST.
In the results obtained without using LST, depth values were affected by texture and specular light reflections on the colonic surfaces.
The proposed method obtained better depth estimation results regardless of them.
%Shapes of large to small structures on the colonic wall, including the haustral folds and ileocecal valve, were reflected on the depth images.
%Furthermore, the estimated depth values have correlation to the ground truth as shown in Table \ref{tab:result_depthvalues}.
The estimated depth values and the ground truth values had a positive correlation (the correlation coefficient was 0.45) in the experiments using the real endoscopic images with a point depth.
This means the estimated depth images represent the shape of colonic walls.
However, absolute values of the estimated depth were different from the ground truth in Table \ref{tab:result_depthvalues}.
The differences were caused by difference of the real and virtual camera parameters.
The camera parameters of the real colonoscope and virtual camera should be calibrated to reduce differences of how these cameras map a target object to images.
In our method, camera calibration was not performed.
We need to calibrate real colonoscope and virtual camera to improve the results.

We conducted an experiment to evaluate the usefulness of the estimated depth images for automated location identification.
The averaged accuracy of the location identification from images was improved from 69.2\% to 74.1\% by using the estimated depth images obtained by using the proposed method.
We obtained the highest accuracy among the depth estimation methods shown in Table \ref{tab:result_cnn}.
The results indicate the LST and ME loss proposed in this paper contribute in obtaining high quality depth estimation results.
The shape information of objects in scenes is quite important for understanding scenes.
However, the sizes of objects are difficult to understand from 2D images because distance information from the camera to the object has lost.
The estimated depth images contributed to recover the distance information and improved the automated location identification accuracy.
%The depth images are thus helpful in understanding objects accurately.
%In our experiment, we input both colonoscopic and estimated depth images to the CNN.
%The CNN can automatically use the depth information to solve the classification  task.
We showed a typical result as one example of using the proposed depth estimation method.
In practice, the proposed method can be applied to many automated endoscopic tasks involving scene understanding.

The depth images can be used to improve automated navigation, tracking, scene understanding, lesion detection, and quantitative analysis of lesions.
By using the depth images, these methods can utilize not only color information but also 3D shape information on the surface of organs or surgical tools.
The proposed method shows quite promising results for enhancing endoscopic diagnosis/treatment assistance.
Our method can estimate depths from monocular single-shot images.
This is an important result to extend our application to stereo cameras and time-series images.

An important clinical application of the depth estimation is size measurement of a colonic polyp during a colon inspection.
Colonic polyps larger than 5mm should be confirmed by a physician whether they are benign or malignant.
Our method provided accurate depth estimation results when a measurement target was close to the colonoscope camera.
Our method can be used to measure a colonic polyp size when a physician finds a suspicious region during an inspection.
To improve clinical value of our method, depth estimation accuracy of targets that are distant from the colonoscope camera should be improved.

\section{Conclusions}

We proposed a depth estimation method from monocular single-shot endoscopic images.
A domain adaptation technique was used to translate a real endoscopic image into a Lambertian surface domain.
The translated image was processed by the depth estimation FCN.
The FCN has a DenseNet-based encoder-decoder structure.
The FCN was trained using the multi-scale edge loss.
In the experiment using real endoscopic images with a point depth, the estimated depth values had positive correlation with the ground truth values.
%The correlation coefficient of them was 0.98.
%In the experiment using real endoscopic images with a point depth, the proposed method estimated accurate depth values.
Also, in the experiment using the location identification CNN, use of the estimated depth image resulted in improvement of averaged identification accuracy, from 69.2\% to 74.1\%.
%Future work includes depth estimation from stereo or time-series images and applications to scene-understanding of other types of images.
Our future work will include investigation of network structures for domain adaptation or depth estimation and applications to scene understanding of other types of images.

\section*{Disclosure statement}
The authors report there are no competing interests to declare.

\section*{Funding} 
Parts of this research were supported by the AMED Grant Numbers 18lk1010028s0401, JP19lk1010036, JP20lk1010036, the MEXT/JSPS KAKENHI Grant Numbers 26108006, 17H00867, 17K20099, the JST CREST Grant Number JPMJCR20D5, and the JSPS Bilateral International Collaboration Grants.

\end{document}